\documentclass[12pt,preprint,aas_macros]{aastex}

\def\PsfigVersion{1.10}
\def\setDriver{\DvipsDriver} % \DvipsDriver or \OzTeXDriver
\ifx\undefined\psfig\else\endinput\fi
\let\LaTeXAtSign=\@
\let\@=\relax
\edef\psfigRestoreAt{\catcode`\@=\number\catcode`@\relax}
\catcode`\@=11\relax
\newwrite\@unused
\def\ps@typeout#1{{\let\protect\string\immediate\write\@unused{#1}}}

\def\DvipsDriver{
	\ps@typeout{psfig/tex \PsfigVersion -dvips}
\def\PsfigSpecials{\DvipsSpecials} 	\def\ps@dir{/}
\def\ps@predir{} }
\def\OzTeXDriver{
	\ps@typeout{psfig/tex \PsfigVersion -oztex}
	\def\PsfigSpecials{\OzTeXSpecials}
	\def\ps@dir{:}
	\def\ps@predir{:}
	\catcode`\^^J=5
}

\def\figurepath{./:}

\def\DoPaths#1{\expandafter\EachPath#1\stoplist}
\def\leer{}
\def\EachPath#1:#2\stoplist{% #1 part of the list (delimiter :)
  \ExistsFile{#1}{\SearchedFile}
  \ifx#2\leer
  \else
    \expandafter\EachPath#2\stoplist
  \fi}
\def\ps@dir{/}
\def\ExistsFile#1#2{%
   \openin1=\ps@predir#1\ps@dir#2
   \ifeof1
       \closein1
   \else
       \closein1
        \ifx\ps@founddir\leer
           \edef\ps@founddir{#1}
        \fi
   \fi}
\def\get@dir#1{%
  \def\ps@founddir{}
  \def\SearchedFile{#1}
  \DoPaths\figurepath
}
\def\@nnil{\@nil}
\def\@empty{}
\def\@psdonoop#1\@@#2#3{}
\def\@psdo#1:=#2\do#3{\edef\@psdotmp{#2}\ifx\@psdotmp\@empty \else
    \expandafter\@psdoloop#2,\@nil,\@nil\@@#1{#3}\fi}
\def\@psdoloop#1,#2,#3\@@#4#5{\def#4{#1}\ifx #4\@nnil \else
       #5\def#4{#2}\ifx #4\@nnil \else#5\@ipsdoloop #3\@@#4{#5}\fi\fi}
\def\@ipsdoloop#1,#2\@@#3#4{\def#3{#1}\ifx #3\@nnil 
       \let\@nextwhile=\@psdonoop \else
      #4\relax\let\@nextwhile=\@ipsdoloop\fi\@nextwhile#2\@@#3{#4}}
\def\@tpsdo#1:=#2\do#3{\xdef\@psdotmp{#2}\ifx\@psdotmp\@empty \else
    \@tpsdoloop#2\@nil\@nil\@@#1{#3}\fi}
\def\@tpsdoloop#1#2\@@#3#4{\def#3{#1}\ifx #3\@nnil 
       \let\@nextwhile=\@psdonoop \else
      #4\relax\let\@nextwhile=\@tpsdoloop\fi\@nextwhile#2\@@#3{#4}}
\ifx\undefined\fbox
\newdimen\fboxrule
\newdimen\fboxsep
\newdimen\ps@tempdima
\newbox\ps@tempboxa
\long\def\fbox#1{\leavevmode\setbox\ps@tempboxa\hbox{#1}\ps@tempdima\fboxrule
    \advance\ps@tempdima \fboxsep \advance\ps@tempdima \dp\ps@tempboxa
   \hbox{\lower \ps@tempdima\hbox
  {\vbox{\hrule height \fboxrule
          \hbox{\vrule width \fboxrule \hskip\fboxsep
          \vbox{\vskip\fboxsep \box\ps@tempboxa\vskip\fboxsep}\hskip 
                 \fboxsep\vrule width \fboxrule}
                 \hrule height \fboxrule}}}}
\fi
\newread\ps@stream
\newif\ifnot@eof       % continue looking for the bounding box?
\newif\if@noisy        % report what you're making?
\newif\if@atend        % %%BoundingBox: has (at end) specification
\newif\if@psfile       % does this look like a PostScript file?
{\catcode`\%=12\global\gdef\epsf@start{%!}}
\def\epsf@PS{PS}
\def\epsf@getbb#1{%
\openin\ps@stream=\ps@predir#1
\ifeof\ps@stream\ps@typeout{Error, File #1 not found}\else
   {\not@eoftrue \chardef\other=12
    \def\do##1{\catcode`##1=\other}\dospecials \catcode`\ =10
    \loop
       \if@psfile
	  \read\ps@stream to \epsf@fileline
       \else{
	  \obeyspaces
          \read\ps@stream to \epsf@tmp\global\let\epsf@fileline\epsf@tmp}
       \fi
       \ifeof\ps@stream\not@eoffalse\else
       \if@psfile\else
       \expandafter\epsf@test\epsf@fileline:. \\%
       \fi
          \expandafter\epsf@aux\epsf@fileline:. \\%
       \fi
   \ifnot@eof\repeat
   }\closein\ps@stream\fi}%
\long\def\epsf@test#1#2#3:#4\\{\def\epsf@testit{#1#2}
			\ifx\epsf@testit\epsf@start\else
\ps@typeout{Warning! File does not start with `\epsf@start'.  It may not be a PostScript file.}
			\fi
			\@psfiletrue} % don't test after 1st line
{\catcode`\%=12\global\let\epsf@percent=%\global\def\epsf@bblit{%BoundingBox}}
\long\def\epsf@aux#1#2:#3\\{\ifx#1\epsf@percent
   \def\epsf@testit{#2}\ifx\epsf@testit\epsf@bblit
	\@atendfalse
        \epsf@atend #3 . \\%
	\if@atend	
	   \if@verbose{
		\ps@typeout{psfig: found `(atend)'; continuing search}
	   }\fi
        \else
        \epsf@grab #3 . . . \\%
        \not@eoffalse
        \global\no@bbfalse
        \fi
   \fi\fi}%
\def\epsf@grab #1 #2 #3 #4 #5\\{%
   \global\def\epsf@llx{#1}\ifx\epsf@llx\empty
      \epsf@grab #2 #3 #4 #5 .\\\else
   \global\def\epsf@lly{#2}%
   \global\def\epsf@urx{#3}\global\def\epsf@ury{#4}\fi}%
\def\epsf@atendlit{(atend)} 
\def\epsf@atend #1 #2 #3\\{%
   \def\epsf@tmp{#1}\ifx\epsf@tmp\empty
      \epsf@atend #2 #3 .\\\else
   \ifx\epsf@tmp\epsf@atendlit\@atendtrue\fi\fi}

\chardef\psletter = 11 % won't conflict with \begin{letter} now...
\chardef\other = 12

\newif \ifdebug %%% turn me on to see TeX hard at work ...
\newif\ifc@mpute %%% don't need to compute some values
\c@mputetrue % but assume that we do

\let\then = \relax
\def\r@dian{pt }
\let\r@dians = \r@dian
\let\dimensionless@nit = \r@dian
\let\dimensionless@nits = \dimensionless@nit
\def\internal@nit{sp }
\let\internal@nits = \internal@nit
\newif\ifstillc@nverging
\def \Mess@ge #1{\ifdebug \then \message {#1} \fi}

{ %%% Things that need abnormal catcodes %%%
	\catcode `\@ = \psletter
	\gdef \nodimen {\expandafter \n@dimen \the \dimen}
	\gdef \term #1 #2 #3%
	       {\edef \t@ {\the #1}%%% freeze parameter 1 (count, by value)
		\edef \t@@ {\expandafter \n@dimen \the #2\r@dian}%
		\t@rm {\t@} {\t@@} {#3}%
	       }
	\gdef \t@rm #1 #2 #3%
	       {{%
		\count 0 = 0
		\dimen 0 = 1 \dimensionless@nit
		\dimen 2 = #2\relax
		\Mess@ge {Calculating term #1 of \nodimen 2}%
		\loop
		\ifnum	\count 0 < #1
		\then	\advance \count 0 by 1
			\Mess@ge {Iteration \the \count 0 \space}%
			\Multiply \dimen 0 by {\dimen 2}%
			\Mess@ge {After multiplication, term = \nodimen 0}%
			\Divide \dimen 0 by {\count 0}%
			\Mess@ge {After division, term = \nodimen 0}%
		\repeat
		\Mess@ge {Final value for term #1 of 
				\nodimen 2 \space is \nodimen 0}%
		\xdef \Term {#3 = \nodimen 0 \r@dians}%
		\aftergroup \Term
	       }}
	\catcode `\p = \other
	\catcode `\t = \other
	\gdef \n@dimen #1pt{#1} %%% throw away the ``pt''
}

\def \Divide #1by #2{\divide #1 by #2} %%% just a synonym

\def \Multiply #1by #2%%% allows division of a dimen by a dimen
       {{%%% should really freeze parameter 2 (dimen, passed by value)
	\count 0 = #1\relax
	\count 2 = #2\relax
	\count 4 = 65536
	\Mess@ge {Before scaling, count 0 = \the \count 0 \space and
			count 2 = \the \count 2}%
	\ifnum	\count 0 > 32767 %%% do our best to avoid overflow
	\then	\divide \count 0 by 4
		\divide \count 4 by 4
	\else	\ifnum	\count 0 < -32767
		\then	\divide \count 0 by 4
			\divide \count 4 by 4
		\else
		\fi
	\fi
	\ifnum	\count 2 > 32767 %%% while retaining reasonable accuracy
	\then	\divide \count 2 by 4
		\divide \count 4 by 4
	\else	\ifnum	\count 2 < -32767
		\then	\divide \count 2 by 4
			\divide \count 4 by 4
		\else
		\fi
	\fi
	\multiply \count 0 by \count 2
	\divide \count 0 by \count 4
	\xdef \product {#1 = \the \count 0 \internal@nits}%
	\aftergroup \product
       }}

\def\r@duce{\ifdim\dimen0 > 90\r@dian \then   % sin(x+90) = sin(180-x)
		\multiply\dimen0 by -1
		\advance\dimen0 by 180\r@dian
		\r@duce
	    \else \ifdim\dimen0 < -90\r@dian \then  % sin(-x) = sin(360+x)
		\advance\dimen0 by 360\r@dian
		\r@duce
		\fi
	    \fi}

\def\Sine#1%
       {{%
	\dimen 0 = #1 \r@dian
	\r@duce
	\ifdim\dimen0 = -90\r@dian \then
	   \dimen4 = -1\r@dian
	   \c@mputefalse
	\fi
	\ifdim\dimen0 = 90\r@dian \then
	   \dimen4 = 1\r@dian
	   \c@mputefalse
	\fi
	\ifdim\dimen0 = 0\r@dian \then
	   \dimen4 = 0\r@dian
	   \c@mputefalse
	\fi
	\ifc@mpute \then
		\divide\dimen0 by 180
		\dimen0=3.141592654\dimen0
		\dimen 2 = 3.1415926535897963\r@dian %%% a well-known constant
		\divide\dimen 2 by 2 %%% we only deal with -pi/2 : pi/2
		\Mess@ge {Sin: calculating Sin of \nodimen 0}%
		\count 0 = 1 %%% see power-series expansion for sine
		\dimen 2 = 1 \r@dian %%% ditto
		\dimen 4 = 0 \r@dian %%% ditto
		\loop
			\ifnum	\dimen 2 = 0 %%% then we've done
			\then	\stillc@nvergingfalse 
			\else	\stillc@nvergingtrue
			\fi
			\ifstillc@nverging %%% then calculate next term
			\then	\term {\count 0} {\dimen 0} {\dimen 2}%
				\advance \count 0 by 2
				\count 2 = \count 0
				\divide \count 2 by 2
				\ifodd	\count 2 %%% signs alternate
				\then	\advance \dimen 4 by \dimen 2
				\else	\advance \dimen 4 by -\dimen 2
				\fi
		\repeat
	\fi		
			\xdef \sine {\nodimen 4}%
       }}

\def\Cosine#1{\ifx\sine\UnDefined\edef\Savesine{\relax}\else
		             \edef\Savesine{\sine}\fi
	{\dimen0=#1\r@dian\advance\dimen0 by 90\r@dian
	 \Sine{\nodimen 0}
	 \xdef\cosine{\sine}
	 \xdef\sine{\Savesine}}}	      
\def\psdraft{
	\def\@psdraft{0}
}
\def\psfull{
	\def\@psdraft{100}
}

\psfull

\newif\if@scalefirst
\def\psscalefirst{\@scalefirsttrue}
\def\psrotatefirst{\@scalefirstfalse}
\psrotatefirst

\newif\if@draftbox
\def\psnodraftbox{
	\@draftboxfalse
}
\def\psdraftbox{
	\@draftboxtrue
}
\@draftboxtrue

\newif\if@prologfile
\newif\if@postlogfile
\def\pssilent{
	\@noisyfalse
}
\def\psnoisy{
	\@noisytrue
}
\psnoisy
\newif\if@bbllx
\newif\if@bblly
\newif\if@bburx
\newif\if@bbury
\newif\if@height
\newif\if@width
\newif\if@rheight
\newif\if@rwidth
\newif\if@angle
\newif\if@clip
\newif\if@verbose
\def\@p@@sclip#1{\@cliptrue}
\newif\if@decmpr
\def\@p@@sfigure#1{\def\@p@sfile{null}\def\@p@sbbfile{null}\@decmprfalse
   \openin1=\ps@predir#1
   \ifeof1
	\closein1
	\get@dir{#1}
	\ifx\ps@founddir\leer
		\openin1=\ps@predir#1.bb
		\ifeof1
			\closein1
			\get@dir{#1.bb}
			\ifx\ps@founddir\leer
				\ps@typeout{Can't find #1 in \figurepath}
			\else
				\@decmprtrue
				\def\@p@sfile{\ps@founddir\ps@dir#1}
				\def\@p@sbbfile{\ps@founddir\ps@dir#1.bb}
			\fi
		\else
			\closein1
			\@decmprtrue
			\def\@p@sfile{#1}
			\def\@p@sbbfile{#1.bb}
		\fi
	\else
		\def\@p@sfile{\ps@founddir\ps@dir#1}
		\def\@p@sbbfile{\ps@founddir\ps@dir#1}
	\fi
   \else
	\closein1
	\def\@p@sfile{#1}
	\def\@p@sbbfile{#1}
   \fi
}
\def\@p@@sfile#1{\@p@@sfigure{#1}}
\def\@p@@sbbllx#1{
		\@bbllxtrue
		\dimen100=#1
		\edef\@p@sbbllx{\number\dimen100}
}
\def\@p@@sbblly#1{
		\@bbllytrue
		\dimen100=#1
		\edef\@p@sbblly{\number\dimen100}
}
\def\@p@@sbburx#1{
		\@bburxtrue
		\dimen100=#1
		\edef\@p@sbburx{\number\dimen100}
}
\def\@p@@sbbury#1{
		\@bburytrue
		\dimen100=#1
		\edef\@p@sbbury{\number\dimen100}
}
\def\@p@@sheight#1{
		\@heighttrue
		\dimen100=#1
   		\edef\@p@sheight{\number\dimen100}
}
\def\@p@@swidth#1{
		\@widthtrue
		\dimen100=#1
		\edef\@p@swidth{\number\dimen100}
}
\def\@p@@srheight#1{
		\@rheighttrue
		\dimen100=#1
		\edef\@p@srheight{\number\dimen100}
}
\def\@p@@srwidth#1{
		\@rwidthtrue
		\dimen100=#1
		\edef\@p@srwidth{\number\dimen100}
}
\def\@p@@sangle#1{
		\@angletrue
		\edef\@p@sangle{#1} %\number\dimen100}
}
\def\@p@@ssilent#1{ 
		\@verbosefalse
}
\def\@p@@sprolog#1{\@prologfiletrue\def\@prologfileval{#1}}
\def\@p@@spostlog#1{\@postlogfiletrue\def\@postlogfileval{#1}}
\def\@cs@name#1{\csname #1\endcsname}
\def\@setparms#1=#2,{\@cs@name{@p@@s#1}{#2}}
\def\ps@init@parms{
		\@bbllxfalse \@bbllyfalse
		\@bburxfalse \@bburyfalse
		\@heightfalse \@widthfalse
		\@rheightfalse \@rwidthfalse
		\def\@p@sbbllx{}\def\@p@sbblly{}
		\def\@p@sbburx{}\def\@p@sbbury{}
		\def\@p@sheight{}\def\@p@swidth{}
		\def\@p@srheight{}\def\@p@srwidth{}
		\def\@p@sangle{0}
		\def\@p@sfile{} \def\@p@sbbfile{}
		\def\@p@scost{10}
		\def\@sc{}
		\@prologfilefalse
		\@postlogfilefalse
		\@clipfalse
		\if@noisy
			\@verbosetrue
		\else
			\@verbosefalse
		\fi
}
\def\parse@ps@parms#1{
	 	\@psdo\@psfiga:=#1\do
		   {\expandafter\@setparms\@psfiga,}}
\newif\ifno@bb
\def\bb@missing{
	\if@verbose{
		\ps@typeout{psfig: searching \@p@sbbfile \space  for bounding box}
	}\fi
	\no@bbtrue
	\epsf@getbb{\@p@sbbfile}
        \ifno@bb \else \bb@cull\epsf@llx\epsf@lly\epsf@urx\epsf@ury\fi
}	
\def\bb@cull#1#2#3#4{
	\dimen100=#1 bp\edef\@p@sbbllx{\number\dimen100}
	\dimen100=#2 bp\edef\@p@sbblly{\number\dimen100}
	\dimen100=#3 bp\edef\@p@sbburx{\number\dimen100}
	\dimen100=#4 bp\edef\@p@sbbury{\number\dimen100}
	\no@bbfalse
}
\newdimen\p@intvaluex
\newdimen\p@intvaluey
\def\rotate@#1#2{{\dimen0=#1 sp\dimen1=#2 sp
		  \global\p@intvaluex=\cosine\dimen0
		  \dimen3=\sine\dimen1
		  \global\advance\p@intvaluex by -\dimen3
		  \global\p@intvaluey=\sine\dimen0
		  \dimen3=\cosine\dimen1
		  \global\advance\p@intvaluey by \dimen3
		  }}
\def\compute@bb{
		\no@bbfalse
		\if@bbllx \else \no@bbtrue \fi
		\if@bblly \else \no@bbtrue \fi
		\if@bburx \else \no@bbtrue \fi
		\if@bbury \else \no@bbtrue \fi
		\ifno@bb \bb@missing \fi
		\ifno@bb \ps@typeout{FATAL ERROR: no bb supplied or found}
			\no-bb-error
		\fi
		\count203=\@p@sbburx
		\count204=\@p@sbbury
		\advance\count203 by -\@p@sbbllx
		\advance\count204 by -\@p@sbblly
		\edef\ps@bbw{\number\count203}
		\edef\ps@bbh{\number\count204}
		\if@angle 
			\Sine{\@p@sangle}\Cosine{\@p@sangle}
	        	{\dimen100=\maxdimen\xdef\r@p@sbbllx{\number\dimen100}
					    \xdef\r@p@sbblly{\number\dimen100}
			                    \xdef\r@p@sbburx{-\number\dimen100}
					    \xdef\r@p@sbbury{-\number\dimen100}}
                        \def\minmaxtest{
			   \ifnum\number\p@intvaluex<\r@p@sbbllx
			      \xdef\r@p@sbbllx{\number\p@intvaluex}\fi
			   \ifnum\number\p@intvaluex>\r@p@sbburx
			      \xdef\r@p@sbburx{\number\p@intvaluex}\fi
			   \ifnum\number\p@intvaluey<\r@p@sbblly
			      \xdef\r@p@sbblly{\number\p@intvaluey}\fi
			   \ifnum\number\p@intvaluey>\r@p@sbbury
			      \xdef\r@p@sbbury{\number\p@intvaluey}\fi
			   }
			\rotate@{\@p@sbbllx}{\@p@sbblly}
			\minmaxtest
			\rotate@{\@p@sbbllx}{\@p@sbbury}
			\minmaxtest
			\rotate@{\@p@sbburx}{\@p@sbblly}
			\minmaxtest
			\rotate@{\@p@sbburx}{\@p@sbbury}
			\minmaxtest
			\edef\@p@sbbllx{\r@p@sbbllx}\edef\@p@sbblly{\r@p@sbblly}
			\edef\@p@sbburx{\r@p@sbburx}\edef\@p@sbbury{\r@p@sbbury}
		\fi
		\count203=\@p@sbburx
		\count204=\@p@sbbury
		\advance\count203 by -\@p@sbbllx
		\advance\count204 by -\@p@sbblly
		\edef\@bbw{\number\count203}
		\edef\@bbh{\number\count204}
}
\def\in@hundreds#1#2#3{\count240=#2 \count241=#3
		     \count100=\count240	% 100 is first digit #2/#3
		     \divide\count100 by \count241
		     \count101=\count100
		     \multiply\count101 by \count241
		     \advance\count240 by -\count101
		     \multiply\count240 by 10
		     \count101=\count240	%101 is second digit of #2/#3
		     \divide\count101 by \count241
		     \count102=\count101
		     \multiply\count102 by \count241
		     \advance\count240 by -\count102
		     \multiply\count240 by 10
		     \count102=\count240	% 102 is the third digit
		     \divide\count102 by \count241
		     \count200=#1\count205=0
		     \count201=\count200
			\multiply\count201 by \count100
		 	\advance\count205 by \count201
		     \count201=\count200
			\divide\count201 by 10
			\multiply\count201 by \count101
			\advance\count205 by \count201
		     \count201=\count200
			\divide\count201 by 100
			\multiply\count201 by \count102
			\advance\count205 by \count201
		     \edef\@result{\number\count205}
}
\def\compute@wfromh{
		\in@hundreds{\@p@sheight}{\@bbw}{\@bbh}
		\edef\@p@swidth{\@result}
}
\def\compute@hfromw{
	        \in@hundreds{\@p@swidth}{\@bbh}{\@bbw}
		\edef\@p@sheight{\@result}
}
\def\compute@handw{
		\if@height 
			\if@width
			\else
				\compute@wfromh
			\fi
		\else 
			\if@width
				\compute@hfromw
			\else
				\edef\@p@sheight{\@bbh}
				\edef\@p@swidth{\@bbw}
			\fi
		\fi
}
\def\compute@resv{
		\if@rheight \else \edef\@p@srheight{\@p@sheight} \fi
		\if@rwidth \else \edef\@p@srwidth{\@p@swidth} \fi
}
\def\compute@sizes{
	\compute@bb
	\if@scalefirst\if@angle
	\if@width
	   \in@hundreds{\@p@swidth}{\@bbw}{\ps@bbw}
	   \edef\@p@swidth{\@result}
	\fi
	\if@height
	   \in@hundreds{\@p@sheight}{\@bbh}{\ps@bbh}
	   \edef\@p@sheight{\@result}
	\fi
	\fi\fi
	\compute@handw
	\compute@resv}
\def\OzTeXSpecials{
	\special{empty.ps /@isp {true} def}
	\special{empty.ps \@p@swidth \space \@p@sheight \space
			\@p@sbbllx \space \@p@sbblly \space
			\@p@sbburx \space \@p@sbbury \space
			startTexFig \space }
	\if@clip{
		\if@verbose{
			\ps@typeout{(clip)}
		}\fi
		\special{empty.ps doclip \space }
	}\fi
	\if@angle{
		\if@verbose{
			\ps@typeout{(rotate)}
		}\fi
		\special {empty.ps \@p@sangle \space rotate \space} 
	}\fi
	\if@prologfile
	    \special{\@prologfileval \space } \fi
	\if@decmpr{
		\if@verbose{
			\ps@typeout{psfig: Compression not available
			in OzTeX version \space }
		}\fi
	}\else{
		\if@verbose{
			\ps@typeout{psfig: including \@p@sfile \space }
		}\fi
		\special{epsf=\ps@predir\@p@sfile \space }
	}\fi
	\if@postlogfile
	    \special{\@postlogfileval \space } \fi
	\special{empty.ps /@isp {false} def}
}
\def\DvipsSpecials{
	\special{ps::[begin] 	\@p@swidth \space \@p@sheight \space
			\@p@sbbllx \space \@p@sbblly \space
			\@p@sbburx \space \@p@sbbury \space
			startTexFig \space }
	\if@clip{
		\if@verbose{
			\ps@typeout{(clip)}
		}\fi
		\special{ps:: doclip \space }
	}\fi
	\if@angle
		\if@verbose{
			\ps@typeout{(clip)}
		}\fi
		\special {ps:: \@p@sangle \space rotate \space} 
	\fi
	\if@prologfile
	    \special{ps: plotfile \@prologfileval \space } \fi
	\if@decmpr{
		\if@verbose{
			\ps@typeout{psfig: including \@p@sfile.Z \space }
		}\fi
		\special{ps: plotfile "`zcat \@p@sfile.Z" \space }
	}\else{
		\if@verbose{
			\ps@typeout{psfig: including \@p@sfile \space }
		}\fi
		\special{ps: plotfile \@p@sfile \space }
	}\fi
	\if@postlogfile
	    \special{ps: plotfile \@postlogfileval \space } \fi
	\special{ps::[end] endTexFig \space }
}
\def\psfig#1{\vbox {
	\ps@init@parms
	\parse@ps@parms{#1}
	\compute@sizes
	\ifnum\@p@scost<\@psdraft{
		\PsfigSpecials 
		\vbox to \@p@srheight sp{
			\hbox to \@p@srwidth sp{
				\hss
			}
		\vss
		}
	}\else{
		\if@draftbox{		
			\hbox{\fbox{\vbox to \@p@srheight sp{
			\vss
			\hbox to \@p@srwidth sp{ \hss 
			 \hss }
			\vss
			}}}
		}\else{
			\vbox to \@p@srheight sp{
			\vss
			\hbox to \@p@srwidth sp{\hss}
			\vss
			}
		}\fi

	}\fi
}}
\psfigRestoreAt
\setDriver
\let\@=\LaTeXAtSign

\newcommand{\rxsj}{1RXS~J141256.0+792204}
\newcommand{\calvera}{Calvera}
\newcommand{\eighteen}{1RXS~J185635.1$-$375433}
\newcommand{\secondsource}{SWIFT~J141055.1+791309}
\newcommand{\thirdsource}{\mbox{CXOU~J141259.43+791958}} % IAU
\newcommand{\hst}{\textit{HST}}
\newcommand{\hstlong}{\textit{Hubble Space Telescope}}
\newcommand{\Swift}{{\em Swift}}
\newcommand{\swift}{\textit{Swift}}
\newcommand{\Chandra}{{\em Chandra}}
\newcommand{\chandra}{\textit{Chandra}}
\newcommand{\chandralong}{\textit{Chandra X-ray Observatory}}
\newcommand{\xmm}{\textit{XMM-Newton}}

\def\deg{\hbox{$^\circ$}}
\def\arcdeg{\hbox{$^\circ$}}
\def\arcmin{\hbox{$^\prime$}}
\def\arcsec{\hbox{$^{\prime\prime}$}}
\def\kmsec{\hbox{${\rm km}\,{\rm s^{-1}}$}}
\def\km{\hbox{${\rm km}$}}
\def\simlt{\mathrel{\hbox{\rlap{\hbox{\lower4pt\hbox{$\sim$}}}\hbox{$<$}}}}
\def\simgt{\mathrel{\hbox{\rlap{\hbox{\lower4pt\hbox{$\sim$}}}\hbox{$>$}}}}

\newcommand{\nh}{\mbox{$N_H$}}
\newcommand{\Mej}{\mbox{$M_{\rm ej}$}}
\newcommand{\Msun}{\mbox{$M_\odot$}}
\newcommand{\nsns}{\mbox{NS-NS}}
\newcommand{\nsbh}{\mbox{NS-BH}}
\newcommand{\xray}{\mbox{X-ray}}
\newcommand{\fix}{\mbox{\textbf{???}}}

\newcommand{\percmsq}{cm$^{-2}$}
\newcommand{\percmcube}{cm$^{-3}$}
\newcommand{\rosat}{{\em ROSAT}}

\newcommand{\usno}{\mbox{USNO-A2.0}}
\newcommand{\usnob}{\mbox{USNO-B1.0}}
\newcommand{\perval}[2]{{#1\mbox{$^{#2}$}}}
\newcommand{\msun}{\mbox{$M_\odot$}}
\newcommand{\rsun}{$R_\odot$}
\newcommand{\persec}{\perval{\rm s}{-1}\/}
\newcommand{\percm}{\mbox{$\cm^{-2}$}}
\newcommand{\peryear}{{{\rm yr}$^{-1}$}}
\newcommand{\ppm}{\mbox{$\pm$}}
\newcommand{\ee}[1]{\mbox{$10^{#1}$}}
\newcommand{\tee}[1]{\mbox{$\times 10^{#1}$}}

\newcommand{\erg}{\mbox{$\rm\,erg$}\/}
\newcommand{\cm}{\mbox{$\rm\,cm$}}
\newcommand{\ksec}{\mbox{$\rm\,ksec$}}
\newcommand{\kpc}{\mbox{$\rm\,kpc$}}
\newcommand{\pc}{\mbox{$\rm\,pc$}}
\newcommand{\fxfopt}{\mbox{$F_{\rm X}/F_{\rm opt}$}}
\newcommand{\fxfv}{\mbox{$F_{\rm X}/F_V$}}
\newcommand{\fxsoft}{\mbox{$F_{\rm X}\mbox{(0.1--2.4 keV)}$}}
\newcommand{\fxstd}{\mbox{$F_{\rm X}\mbox{(2--10 keV)}$}}
\newcommand{\kteff}{\mbox{$kT_{\rm eff}$}}

\newcommand{\ud}[2]{\mbox{$^{+ #1}_{- #2}$}}
\newcommand{\cgsflux}{\erg\,\percm\,\persec}
\newcommand{\cgslum}{\erg\,\persec}
\newcommand{\rbb}{\mbox{$R_{\rm bb}$}}

\begin{document}

%%%%%%%%%%%%%%%%%%%%%%%%%%%%%%%%%%%%%%%%

\title{Discovery of an Isolated Compact Object \\ at High Galactic Latitude}

\author{%
   R.~E.~Rutledge\altaffilmark{1},
   \altaffiltext{1}{Department of Physics, McGill University,
        3600 rue University, Montreal, QC, H3A 2T8, Canada;
        rutledge@physics.mcgill.ca} 
   D.~B.~Fox\altaffilmark{2},
   \& A.~H.~Shevchuk\altaffilmark{2}
   \altaffiltext{2}{Department of Astronomy \& Astrophysics,
        525 Davey Laboratory, Pennsylvania State University, 
        University Park, PA 16802, USA; 
        dfox@astro.psu.edu, ahs148@psu.edu}
}
\received{April 27, 2007}
\slugcomment{ApJ, in press}
\shorttitle{A compact object at high Galactic latitude}
\shortauthors{Rutledge et al.}

%%%%%%%%%%%%%%%%%%%%%%%%%%%%%%%%%%%%%%%%

\begin{abstract}

We report discovery of a compact object at high Galactic latitude.
The object was initially identified as a \rosat\ All-Sky Survey Bright
Source Catalog \xray\ source, \rxsj, statistically likely to possess a
high \xray\ to optical flux ratio.  Further observations using {\em
  Swift}, Gemini-North, and the {\em Chandra} X-ray Observatory
refined the source position and confirmed the absence of any optical
counterpart to an \xray\ to optical flux ratio of $\fxsoft/F_V > 8700$
(3$\sigma$).  Interpretation of \rxsj -- which we have dubbed
\calvera\ -- as a typical X-ray-dim isolated neutron star would place
it at $z\approx 5.1$\,kpc above the Galactic disk -- in the Galactic
halo -- implying that it either has an extreme space velocity
($v_z\simgt 5100$\,\kmsec) or has failed to cool according to
theoretical predictions.  Interpretations as a persistent anomalous
X-ray pulsar, or a ``compact central object'' present conflicts with
these classes' typical properties.  We conclude the properties of
\calvera\ are most consistent with those of a nearby (80 to 260\,pc)
radio pulsar, similar to the radio millisecond pulsars of 47~Tuc, with
further observations required to confirm this classification.  If it
is a millisecond pulsar, it has an X-ray flux equal to the
\xray\ brightest millisecond pulsar (and so is tied for highest flux);
is the closest northern hemisphere millisecond pulsar; and is potentially
the closest known millisecond pulsar in the sky, making it an
interesting target for \xray-study, a radio pulsar timing array, and
{\em LIGO}.
\end{abstract}


\maketitle

%%%%%%%%%%%%%%%%%%%%%%%%%%%%%%%%%%%%%%%%

\section{Introduction}
\label{sec:intro}

During the past decade, seven isolated neutron stars (INSs),
exhibiting no detectable radio pulsar activity, have been discovered
in nearby regions of our Galaxy via \xray\ surveys.  These INSs
exhibit thermal emission spectra with peak energies in the
far-ultraviolet or soft \xray, from atmospheres that are thought to be
amenable to theoretical modeling (e.g., \citealt{lattimer04,pr06}).
As such, they are promising candidates for precision radius
measurements via \xray\ spectroscopy -- measurements that ultimately
aim to constrain theories of the nuclear equation of state.

We are currently pursuing a long-term project to identify new isolated
neutron stars from within the \rosat\ All-Sky Survey Bright Source
Catalog (BSC; \citealt{voges99}) using short $\sim$1000~s observations
with NASA's \swift\ satellite \citep{swift}.  Although the scientific
focus of the \swift\ mission is on gamma-ray burst (GRB) studies, its
autonomous pointing capability, rapid one degree per second slewing,
and joint \xray\ + UV/optical instrumentation make it an excellent
platform for such large-scale surveys \citep{fox04}, as well as for
generic targets of opportunity (TOOs).

We select our targets for \swift\ using statistical techniques of
catalog cross-correlation \citep{rutledge03b}, comparing sources from the
BSC to the \usno, IRAS point-source, and NVSS catalogs, and selecting
sources that are likely, on general grounds, to possess high $L_{\rm
X}/L_{\rm opt/IR/radio}$ (see also \citealt{xid}).  Our candidate
INS sources are then observed with \swift\ on a time-available basis as
``fill-in targets,'' with a low priority that is guaranteed not to
preempt GRB and time-sensitive TOO observations.

Before discussing the results of our survey, we note that the term
``isolated neutron star'' (INS), although it appears generic, has in
fact been adopted to refer exclusively to the seven known non-radio
pulsar, non-magnetar, non supernova-remnant associated neutron stars.
Since we will be selecting targets on the basis of a high \xray\ to
optical flux ratio, $\fxsoft/F_V$, which admits these several types of
compact object into our sample, we find it necessary to define a more
inclusive term: the ``isolated compact object'' or ICO.  The ICO is an
object demonstrated to have a sufficiently high $\fxsoft/F_V$ that its
nature as a compact object is assured, with the only remaining
question being what type of compact object it is -- and an INS being
one such possibility.

In this paper we discuss the first ICO (and candidate INS) to be
identified from our survey, \rxsj.  In \S\ref{sec:obs}, we present our
suite of observations of this source, which demonstrate an \xray\ to
optical flux ratio $\fxsoft/F_V > 8700$, sufficient to establish it as
a compact object.  Given its unique set of properties, we re-christen
the source \calvera, and in \S\ref{sec:discuss}, discuss its
properties in relation to known compact object classes.  Establishing
a preferred interpretation, while challenging in the absence of
further observations, proves to be a useful exercise, casting light on
larger questions related to the nature of the known populations of
isolated compact objects.  Our current state of knowledge and best
interpretation of \calvera\ are then presented in
\S\ref{sec:conclude}.

%%%%%%%%%%%%%%%%%%%%%%%%%%%%%%%%%%%%%%%%

\section{Observations \& Analysis}
\label{sec:obs}

We observed the Bright Source Catalog source \rxsj\ with the \swift\
satellite beginning at 01:25 UT on 25~August 2006.  The single
observation, extending over two satellite orbits, yielded 1.9~ksec of
simultaneous exposure with the satellite's \xray\ Telescope (XRT;
\citealt{xrt}) and Ultraviolet/Optical Telescope (UVOT;
\citealt{uvot}) instruments.  UVOT observations were taken exclusively
through the UVM2 filter, with filter transmission peaking at 220~nm,
and XRT observations were in Photon Counting mode.  We reduced the
data using a custom pipeline derived from official mission data
analysis tools (HEASOFT, v6.2) and CIAO v3.4 \citep{ciao}, with XRT
event grades 0--12 accepted for analysis.

The \swift\ XRT data recover the bright \xray\ source \rxsj, allow an
improved estimate of the source position, and provide a broadband
\xray\ spectrum.  The X-ray source position and its statistical
uncertainty were derived using {\tt wavdetect}.  We extracted 81
source counts from within 30\arcsec\ of the best position, and 51
background counts from a 3.3\arcmin\ radius circle centered
7.3\arcmin\ from the source position.  The expected number of
background counts in the source region is 1.2\ppm0.2, which we neglect
in the forgoing analysis. 

This \xray\ spectrum (Table~\ref{tab:props}) was fit using the
standard products response matrix {\tt swxpc0to12\_20010101v008.rmf}
and an ancillary response matrix produced using the FTOOLS task {\tt
  xrtmkarf}, in XSPEC v12.3.0 \citep{xspec}.  The column density (\nh)
is held fixed at the Galactic value \citep{dickey90}, because the
small number of counts do not permit a strong constraint on \nh; also,
\nh\ is sufficiently low (3\tee{20} \perval{cm}{-2}) that either
doubling \nh\ or setting \nh=0 \perval{cm}{-2} result in an effective
temperature (\kteff) within the best fit 90\% confidence statistical
uncertainty.  

We first attempted to bin the spectrum into four spectral bins,
corresponding to energy ranges 0.3-1.0, 1.0-2.0, 2.0-5.0 and 5.0-10
keV, which contain 44, 28, 2 and 0 counts, respectively.  The number
of counts in these bins is so low, that for two such bins, the usual
assumption for $\chi^2$ statistics, that the error bars can be
approximated as Gaussian, does not hold; thus, with two spectral bins
where this assumption does hold (with number of counts $>$15) and two
spectral parameters to be fit, the observed X-ray spectrum has too low
of signal-to-noise ratio to exclude a blackbody spectral model (or any
other $\geq$2 parameter model). However, we can derive maximumum
likelihood spectral parameters, and their uncertainties; a statistic
which is better for doing so when the number of X-ray counts is low is
the Cash statistic \citep{cash79}, which we use to derive the maximum
likelihood parameter values and their uncertainties.

The INSs are all well-described by thermal (blackbody) spectral
models.  The best-fit blackbody model for \calvera\ gives
\kteff=215\ppm25 eV, and \rbb=7.3\ud{2.6}{1.7} km (d/10\kpc)
(uncertainties given are 90\% confidence).

Alternatively, the best-fit photon power-law slope is 2.8\ppm0.3; a
thermal bremsstrahlung model has best-fit $kT_{\rm
 Bremss}=0.8\ud{0.3}{0.2}\,{\rm keV}$, and the best-fit Raymond-Smith
plasma has a temperature $kT_{\rm R-S}=1.5\ud{0.6}{0.3}$ keV.

The simultaneous UVOT observation reveals no 220~nm counterpart to a
two-sigma limiting magnitude of 21.6~mags, or $1.27 \times 10^{-17}$\,
\cgsflux.

\begin{figure}
\centerline{~\psfig{file=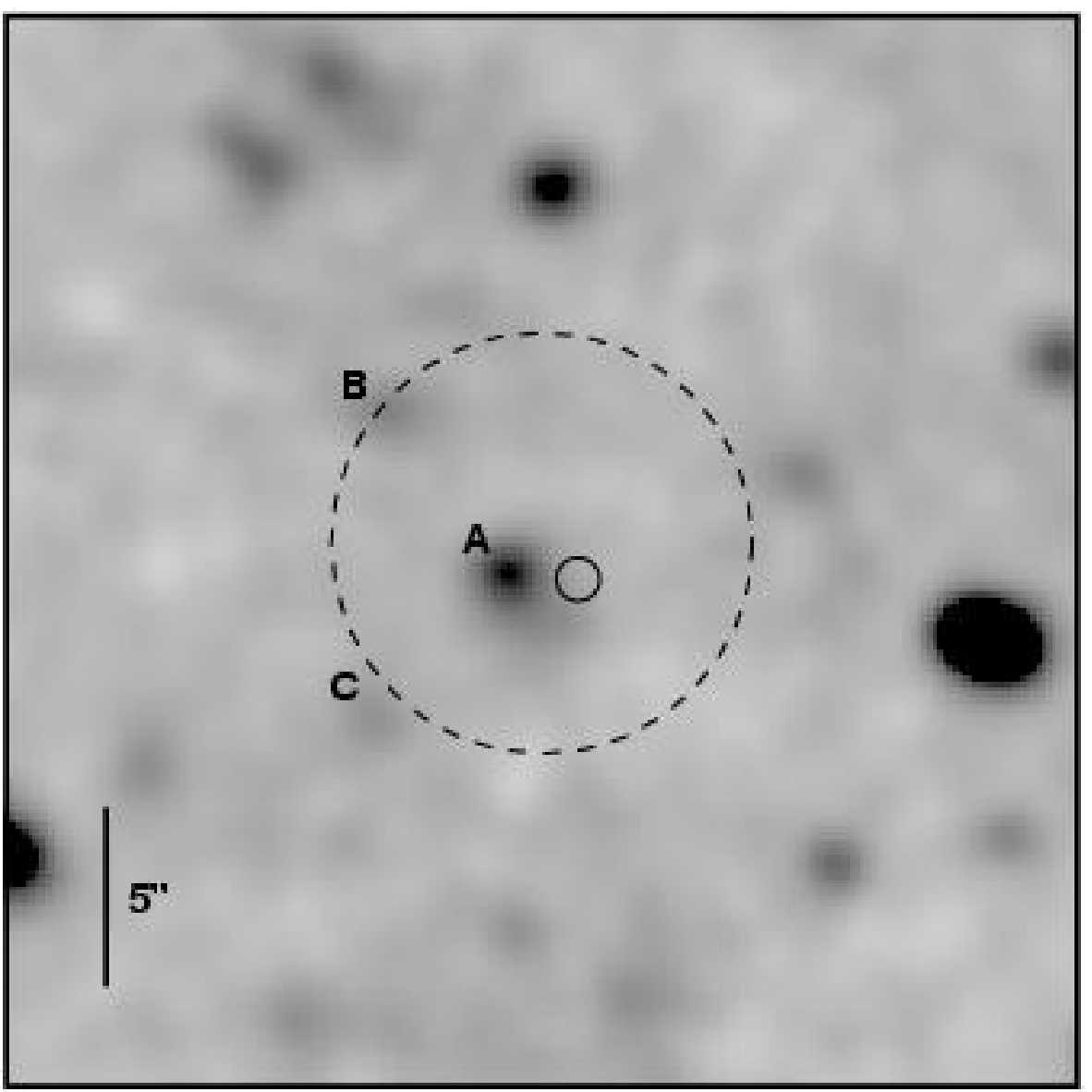,width=5.0in}~}
\bigskip
\caption[]{%
Gemini North + GMOS $g$-band image of the 90\%-confidence \xray\
(\chandra\ HRC-I) localization for \calvera\ (solid circle, radius
0.57\arcsec) , showing the absence of an optical
counterpart to deep limits; the dashed circle (radius 5.9\arcsec)
indicates the 90\%-confidence \swift\ XRT localization. Our
three-sigma limit over the \chandra\ localization is $g>26.3$\,mag,
giving $F_{\rm X}/F_V>8700$ after accounting for extinction.  The
image is smoothed with a PSF-like Gaussian kernel of FWHM 1.2\arcsec\
to reveal faint sources, three of which (A--C) are consistent with the
\swift\ localization.  North is up, East is to the left, and the
image scale is indicated. The nearest point source (``A''), southeast
of \chandra\ localization, has $g=24.8$\,mag and is excluded as a counterpart
with $>3.8\sigma$ confidence.
\label{fig:cxogem}}
\end{figure}

The high \xray\ to UV flux ratio established by the \swift\
observations, combined with limits on the brightness of any optical
(DSS) or near-infrared (2MASS) counterparts from archival surveys, led
us to pursue Director's Discretionary Time observations with the
Gemini telescope.  Our subsequent Gemini North + Gemini Multiobject
Spectrograph (GMOS-N) observations were taken in queue-observing mode
as $4\times 600$\,s images with mean epoch 14:46~UT on 22 Dec 2006.
Data were bias-subtracted, flat-fielded, fringe-subtracted,
registered, and combined with cosmic-ray rejection using Gemini
pipeline software and official calibration products; the photometric
zero-point was derived by reference to ten unsaturated
\mbox{USNO-B1.0} catalog stars in the image.  Our resulting image of
the region of sky around the source reveals three potential optical
counterparts (A--C) with $g$-band magnitudes of $g_1=24.80(8)$ mag,
$g_2=25.64(10)$ mag, and $g_3=25.86(12)$ mag, respectively, where the
uncertainties in the final significant digits are given in
parentheses.  These sources are at J2000 coordinates
$\alpha_1=14:12:56.64$, $\delta_1=+79:22:04.4$;
$\alpha_2=14:12:57.76$, $\delta_2=+79:22:08.8$; and
$\alpha_3=14:12:57.95$, $\delta_3=+79:22:00.2$, respectively, and can
be seen in Fig.~\ref{fig:cxogem}, which shows our final image,
smoothed with a PSF-like Gaussian kernel of FWHM 1.2\arcsec\ to reveal
faint sources.

In order to eliminate the possibility of associating \rxsj\ with these
faint optical sources, we obtained a 2.1$\,$ksec DDT observation with
the {\em Chandra}/HRC-I \citep{hrc1}. A wide-field (imaging) observing
mode, which has degraded fast-timing capabilities, was selected to
include within the field-of-view an unrelated \xray\ point source
(\secondsource; $\approx$ 10\arcmin\ from \calvera), also detected in
our \swift\ XRT observation.  We detected 190 counts within
2\arcsec\ of the best-fit position for \calvera, where only
0.388\ppm0.004 are expected due to background.

Our \texttt{wavdetect} analysis reveals just one other faint
\xray\ source, \thirdsource, at 2\arcmin\ off-axis.
This source lies within the field of view (FOV) of our Gemini
observations, and can be associated with a bright point-like optical
source in this image.  It thus offers the opportunity to make a direct
registration of the \chandra\ and Gemini observations.

We calculate our best estimates for the \chandra\ positions of
\calvera\ and \thirdsource\ using an iterative, weighted-centroiding
algorithm that weights photon positions according to a Gaussian
approximation of the PSF, which we find to be a superior approach to
standard unweighted centroid analyses.  Uncertainties in the resulting
positions are determined by bootstrap Monte Carlo analysis: in each of
10,000 trials, we resample all detected photons within a large window
around the target object, drawing with replacement, and then executing
our centroiding algorithm in full, saving all resulting positions.
Our quoted 90\%-confidence radius corresponds to the minimum-radius
circle enclosing 90\% of the resulting Monte Carlo positions.

We find a \chandra\ position for \thirdsource\ of R.A. 14:12:59.45,
Dec.\ +79:19:58.00 (J2000), with 90\%-confidence uncertainty of
0.48\arcsec, which is distinctly offset from our Gemini position of
R.A. 14:12:59.43, Dec.\ +79:19:58.95 (J2000).  We prefer to quote
positions from the Gemini astrometry, which is registered to the
\mbox{USNO-B1.0} catalog astrometry with precision
0.14\arcsec\ precision (RMS), and note that since our final aim is to
position \calvera\ on the Gemini image, any overall error in the
Gemini astrometry has no impact in this context.  Within 2\arcsec\ of
the best-fit position, we find 8 source counts. \thirdsource\ was not
detected in the SWIFT observation, as its flux (5\tee{-14} \cgsflux,
assuming 1 HRC count = 1.3\tee{-11} erg \perval{cm}{-1}, for photon
powerlaw slope 2.0 and \nh=3\tee{20} \perval{cm}{-2}) was below the
detection limit for the SWIFT observation. 

The registration of \thirdsource\ results in an offset of 0.04\arcsec\
West, 0.95\arcsec\ North ($\approx$0.95\arcsec\ total) to \chandra\
coordinate positions, yielding the position for \calvera\ that we
quote in Table~\ref{tab:props}: R.A. 14:12:55.885, Dec.\ +79:22:04.10
(J2000), with a 90\%-confidence uncertainty of 0.57\arcsec\ that
incorporates the centroiding uncertainty for \calvera\ as well as,
conservatively, the 0.14\arcsec\ precision of the Gemini astrometric
mapping. There are 533 USNO-B1.0 objects with $R>18.6$ within 10
\arcmin\ of \thirdsource; the probability of one lying
$<$0.95\arcsec\ from the \chandra\ position is 1.3\tee{-3}, which we
regard as sufficient to demonstrate the \chandra\ source is associated
with the Gemini source. 

Finally, we note that our $0.95\pm 0.22$ arcsec offset from the
\chandra\ native astrometry is anomalous according to the mission
operations team, with 0.6\arcsec\ radial uncertainty at
90\%-confidence being the expected performance.  However, following
our observations the Chandra \xray\ Center reported discovery of a
0.5\arcsec\ systematic offset in ACIS observations taken after
December 2006, attributed to an altered spacecraft thermal
environment, which may be relevant in this context\footnote{ Chandra
 Electronic Bulletin \# 60; http://cxc.harvard.edu/bulletin/bulletin\_60.html}.

Figure~\ref{fig:cxogem} displays the \xray\ localization for \rxsj\ in
the context of our Gemini $g$-band image.  The nearest point source
(``A''), southeast of the \chandra\ localization, has $g=24.8$\,mag
and is excluded as a counterpart with $>3.8\sigma$ confidence.  We
therefore detect no $g$-band counterpart to a 3$\sigma$ limit of
$g>26.3$\,mag, or $F_g < 0.11$\,$\mu$Jy at 4750\,\AA.  The expected
extinction is $A_g = 0.18$\,mag towards the source, as estimated from
Galactic extinction maps as well as the source's \xray\ spectrum.  As
the $g$ and $V$-band zero points are the same, we find an
extinction-corrected \xray\ to optical flux ratio of $\fxsoft/F_V >
8700$ for \rxsj.

A power density spectrum finds no evidence for periodicity at any
frequency between 4.66\tee{-4} and 1024~Hz with weak limits on \%
root-mean-squared variability of $<$41\%, high compared with detected
periodicities from known INSs \citep{durant06}.  The total observed
HRC-I average countrate is 52 c/s; the nature of the HRC-I wiring
error is such that the absolute timing uncertainty is $\sim$ 0.02
sec\footnote{A technical description of this error and its effect on
the time-tags of individual photons is available at
http://cxc.harvard.edu/cal/Hrc/timing.html}.  Thus, while it is
outside the scope of this paper to fully address the sensitivity of
observations with degraded timing capabilities to a periodic signal,
the $<$41\% r.m.s. variability limit will apply to periodicities in
the range $\sim$0.001-50 Hz, but not to higher frequency
periodicities.

Comparing the spatial distribution of events in the HRC-I focal plane
to the distribution expected for a point source with the spectrum of
\rxsj, using an 85,000-photon ChaRT simulation \citep{ckj+03},
indicates that the source remains unresolved at sub-arcsecond
\chandra\ resolution.  Our 90\%-confidence upper limits on the
fractional contribution from any resolved component are $<$16\% for a
resolved component with Gaussian FWHM of 1~arcsec, and $<$10\% for a
resolved component with Gaussian FWHM $\simgt$ 5~arcsec.

%%%%%%%%%%%%%%%%%%%%%%%%%%%%%%%%%%%%%%%%%%%%%%%%%%

\section{Discussion}
\label{sec:discuss}

The \xray\ to optical flux ratio of \rxsj, $\fxsoft/F_V > 8700$, is
comparable to published limits that have accompanied the discovery of
several INSs:
\eighteen\ (\fxfv$>$7000, \citealt{walter96}),
RX~J1605.3+3249 (\fxfv$\geq$\ee{4}, \citealt{motch99}),
1RXS~J130848.6+212708 (\fxfv$>$12000, \citealt{schwope99}), and 
1RXS J214303.7+065419 (\fxfv$>$\ee{3}, \citealt{zampieri01}).
In all cases the extreme value of \fxfv\ was used to exclude all
non-compact object source classes.  Indeed, all these sources have
subsequently been confirmed as INSs, and no counterexamples -- sources
identified as ICOs on the basis of \fxfv\ and subsequently
demonstrated not to be compact objects -- exist in the literature.
The fact that high \fxfv, sometimes coupled with detection of X-ray
pulsations, is used to identify INSs
\citep{walter96,walter97,haberl97,motch99,schwope99,zampieri01},
and that none has yet been shown to be another class of source, argues
for the robustness of this approach.  We conclude that \rxsj\ is a new
ICO and candidate INS, for which we adopt the name ``\calvera.''

With neither a direct distance measurement nor a proposed association
for \calvera, however -- and without high signal-to-noise \xray\
spectroscopy to confront emission models -- our discussion of the
physical properties of the source must begin by exploring the range of
possible emission models, accepting that both the spectral form and
emitting area are largely unconstrained at present.  We will therefore
discuss each possible model for \calvera\ in turn: isolated neutron
star, magnetar, compact central object, or radio pulsar.

\subsection{Classifying Calvera}
\label{sub:classify}

\begin{figure}
\centerline{~\psfig{file=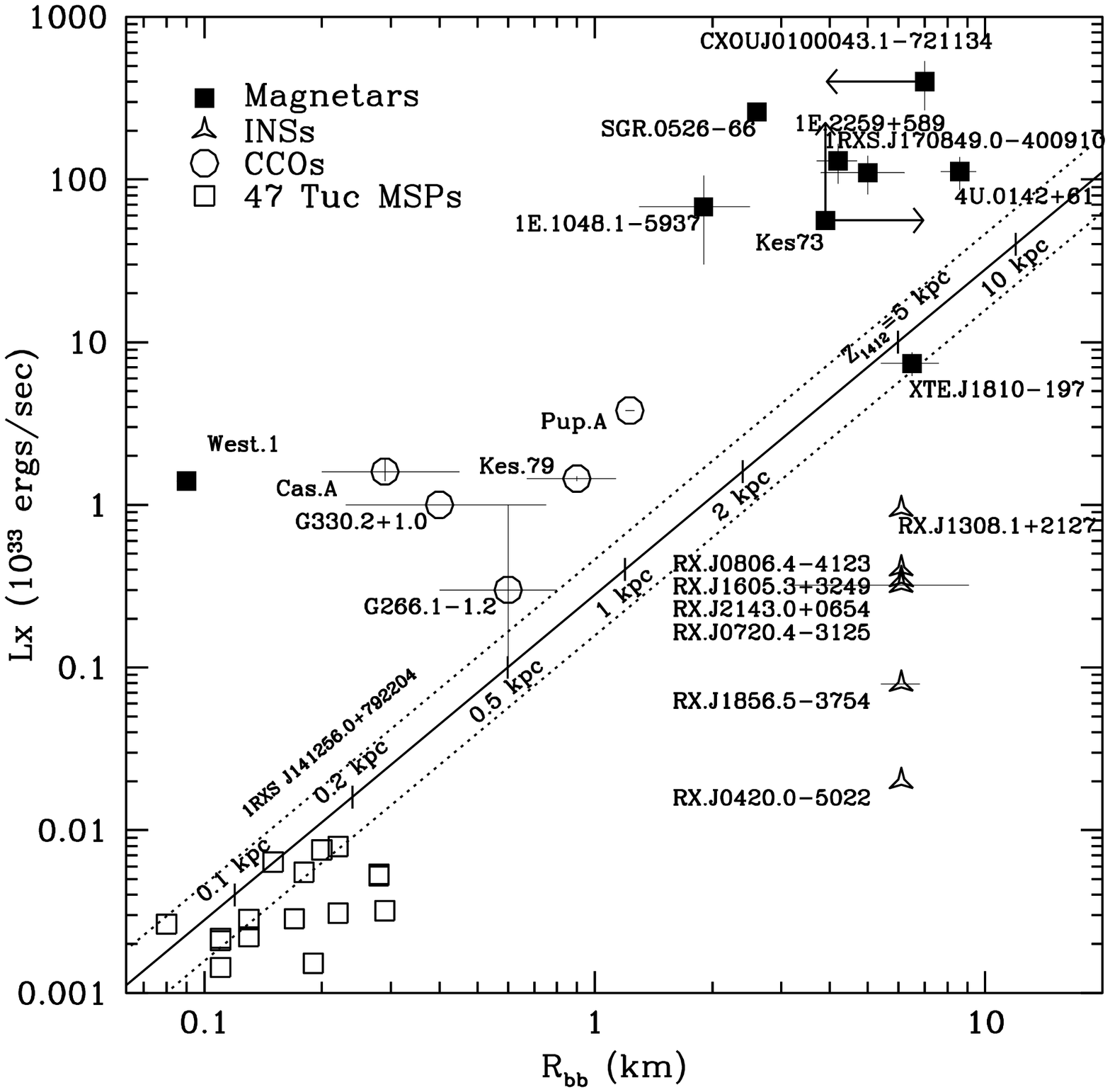,width=5in}~}
\bigskip
\caption[]{%
The blackbody radius \rbb\ and $L_{\rm X}$ values for magnetars (solid
squares), CCOs (open circles), INSs (3-pointed stars), and MSPs in
47~Tuc (open squares).  The distances of the INSs have been normalized
to the parallax distance of \eighteen, assuming all have the same
\rbb; see Table~\ref{tab:dist} and text for details.  The solid line
indicates the best-fit \rbb\ vs. $L_x$, hash-marked with the $z$ for
\calvera; the dotted-lines contain the 90\% uncertainty in \rbb\ for
\calvera. Interpreting \calvera\ as an INS implies that it lies a
vertical distance $z=5.1$\,kpc above the Galactic disk, at a
Galactocentric distance of $R_c = 14.0$\,kpc, well outside the disk
and in the Galactic halo. Interpreting \calvera\ as a persistent
magnetar leads to a paradox, since all other magnetars are found
within a vertical distance $|z|<0.1$\,kpc of the disk, yet a large $z$
is required for \calvera\ to have a luminosity ($L_{\rm X}\approx$
\ee{35} \cgslum) appropriate to the population.  Even with a
luminosity as low as that of the transient magnetar, XTE~J1810-197,
$z>4.9$\,kpc is required.  Interpreting \calvera\ as a CCO -- despite
the absence of any surrounding supernova remnant -- implies that it is
either the faintest member of this population, or lies at a vertical
distance $z\approx 1$ to 5 kpc above the disk. \calvera\ has $L_x$ and
\rbb\ consistent with those of the radio MSPs in 47~Tuc, if its
distance is in the range 80-260 pc.
\label{fig:compare}}
\end{figure}

\begin{figure}
\centerline{~\psfig{file=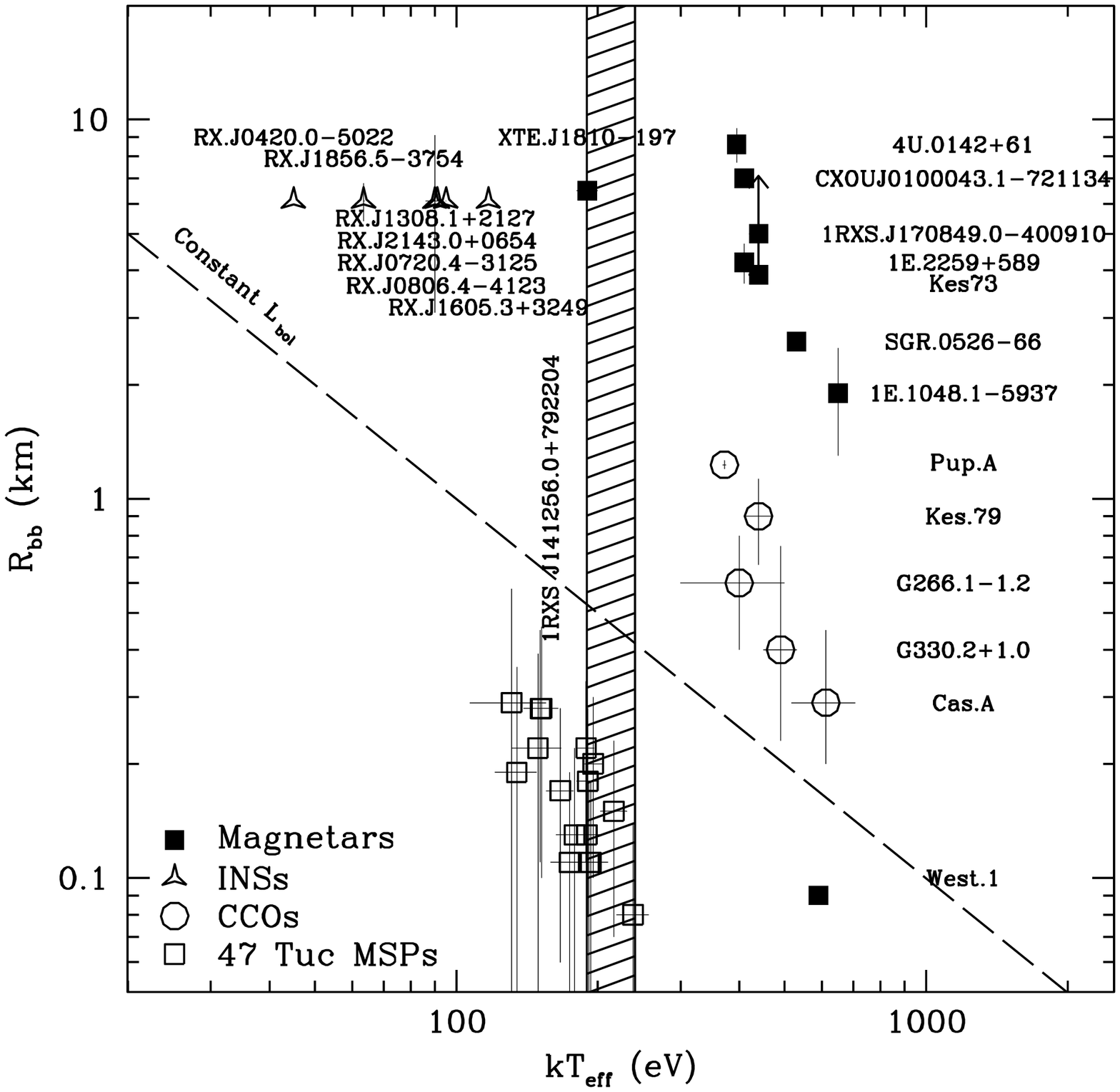,width=5in}~}
\bigskip
\caption[]{%
The effective temperature \kteff\ and blackbody radius \rbb\ for
magnetars, CCOs, INSs, and MSPs in 47~Tuc (see Fig.~\ref{fig:compare}
and text for discussion).  The hashed region indicates the \kteff\
value for \calvera; it covers all values of \rbb\ due to the unknown
distance.  It overlaps \kteff\ for the 47~Tuc MSPs, if \calvera\ is
within a distance range of roughly 80 to 260 pc.
\label{fig:compare2}}
\end{figure}

In this section, we consider four possible compact object
classifications.  In doing so, it will be useful to refer to two
figures.  In Figures~\ref{fig:compare} and \ref{fig:compare2} we
compare the X-ray luminosities, thermal blackbody radii (\rbb) and
effective temperatures (\kteff) for four classes of compact objects --
INSs, magnetars, MSPs and CCOs -- with \calvera.  Here, the
luminosities for the CCOs (Pup A, \citealt{hui06}; G266.1-1.2,
\citealt{pavlov01}; Cas A, \citealt{pavlov00}; G330.2+1.0,
\citealt{park06}) and INSs (from \citealt{haberl05} and references
therein) are the thermal bolometric luminosities.  We use the
bolometric thermal luminosities of the MSPs in 47~Tuc
\citep{bogdanov06} as indicative of those in the field.  Magnetar
luminosities \citep{durant06,muno07} are in the 2--10~keV band
(bolometric luminosities will be greater).

Before attempting to classify \calvera\ within the context of these
two figures, and their corresponding source populations, we make a
general observation.  The CCOs, magnetars and INSs are all thought to
be thermally powered (i.e., due to core and crustal cooling, including
effects of magnetic field decay; \citealt{arras04}).  These classes
occupy different areas of the (\kteff, \rbb) diagram.  The youngest
objects \citep[CCOs, with the object in Westerlund 1 being the notable
exception;][]{muno06} are located in the lower right (high \kteff, low
\rbb), older objects (magnetars) are in the upper right (high \kteff,
high \rbb), and the oldest objects (INSs) are in the upper-left (low
\kteff, high \rbb).  This may suggest a population synthesis model in
which individual objects evolve with time through the (\kteff, \rbb)
diagram.  Although this is speculative -- and we leave detailed
consideration to future work -- for our present purposes it is
sufficient to point out that these three classes of sources, plus the
MSPs of 47~Tuc, occupy different areas of the (\kteff, \rbb) and
(\rbb, $L_X$) diagrams.  The diagrams are therefore a useful means to
compare the properties of \calvera\ to those of each class of compact
object.  

In the following sections, we compare \calvera's properties with the
typical source properties of known classes of compact objects which
have been identified among field \xray\ sources.  The alternative is
that \calvera\ has unique properties -- for example, an \rbb\ or $L_X$
unlike those of previously identified classes, perhaps due to a
different atmospheric composition, magnetic field configuration, or
object size, which we call the ``It Can Be Any Size'' hypothesis
(ICBAS).  The implication of the ICBAS hypothesis is that no statement
about \calvera's distance (and thus luminosity or \rbb) can be made.
The ICBAS hypothesis cannot be falsified in the absence of a distance
measurement; we therefore do not discuss it further.

%%%%%%%%%%%%%%%%%%%%%%%%%%%%%%%%%%%%%%%%%

\subsubsection{Isolated Neutron Star}
\label{sub:ins}

Since the observational approach used to classify \calvera\ is
identical to that of the observationally homogeneous class of INSs --
the so-called Magnificent Seven \citep{haberl06} -- one might expect
that \calvera\ shares observational properties with this class.  As we
show in this section, the properties of \calvera\ diverge from the
known INSs.

Despite deep observations of the seven INSs with \chandra\ and \xmm,
the emergent \xray\ spectra of objects in this class are not well
understood.  The spectrum of the brightest one, \eighteen, is
accurately described by a blackbody in the \xray\ passband
\citep{burwitz03}, but its optical
emission, while also thermal, requires a second blackbody component
with a lower temperature and larger emission area \citep{pons02}. The
only proposed X-ray spectrum which is consistent with both X-ray and
optical spectra is composed of an optically thin layer of magnetized
hydrogen \citep{ho07}.  Other INSs also display thermal spectra in the
\xray\ passband.

\begin{figure}
\centerline{~\psfig{file=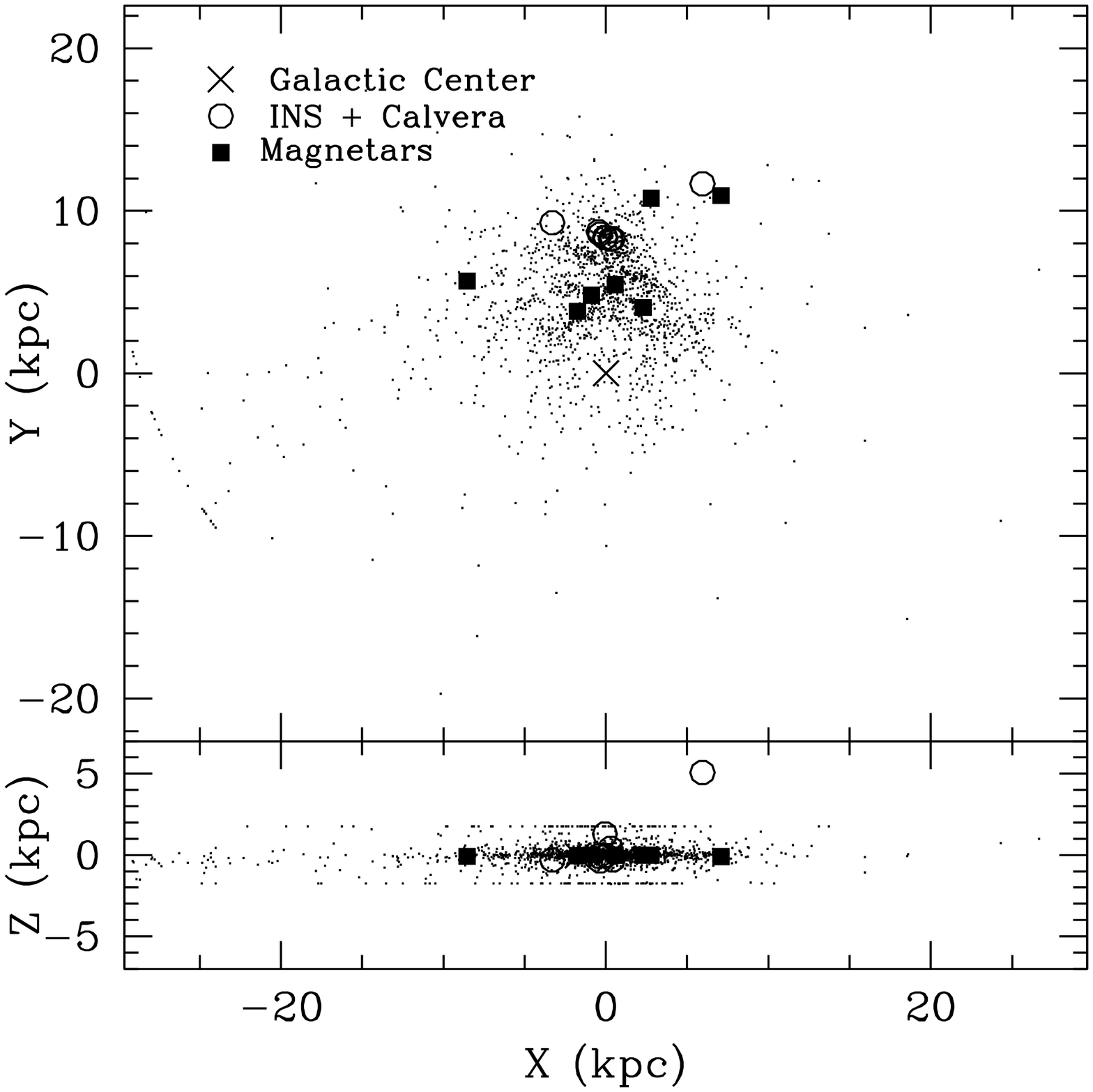,width=5in}~}
\bigskip
\caption[]{%
The galactic distribution of radio pulsars in the ATNF pulsar
catalog (black points; \citealt{manchester05}) -- with objects associated with
globular clusters and known extragalactic sources removed -- magnetars
(solid squares; distances from \citealt{durant06,muno06}), and INSs
and Calvera (assuming INSs and \calvera\ have the same \rbb; open
circles, from Table~\ref{tab:dist}). The implication of interpreting
all INSs as having the same \rbb\ is that they are a spatially
homogeneous population -- local, within the disk of the galaxy.
However, including \calvera\ in this class places it well above the
Galactic disk (bottom panel) at $z\approx5.1\,\kpc$.
\label{fig:dist}}
\end{figure}

Modeling the \xray\ spectra of all INSs as thermal blackbodies of
identical emission area (\rbb=$d\,T_{\rm eff}^{-2}\, \sqrt{{\rm
(B.C.)}\, F_{\rm X} /\sigma}$, where $d$ is source distance, $F_{\rm
X}$ is the 0.1--2.4 keV flux, B.C. is the bolometeric correction,
$\sigma$ is the Stefan-Boltzman constant), we normalize the INS
distances to the parallax distance $d=167\ud{18}{12}\,\pc$ for
\eighteen\ \citealt{kaplan07}, while noting that this recent result is
larger than previous derived distances
\citep{walter02,kaplan02b}. 
The resulting inferred luminosities and three-dimensional locations of
the seven known INSs, and \calvera, are given in Table~\ref{tab:dist},
and displayed in Fig.~\ref{fig:dist}. 

While our approach assumes $\rbb$ is identical for all INSs, this
assumption may not be correct: e.g., compare the range of \rbb\ values
measured for various magnetars (\citealt{durant06}; 2--10 km -- a
factor of 5). It nonetheless is consistent with the newly-measured
parallax distance for the INS RX~J0720.4$-$31225 to the level of
accuracy required for the present comparison (parallax distance
$d=360\ud{170}{90}$\,pc and blackbody distance $d\approx 500$\,pc;
\citealt{kaplan07}).

Calculated in this way, all but one previously-known INS have
distances from the midplane of the Galactic disk $|z|<0.5\,\kpc$ (the
exception is 1RXSJ~130848.6+212708, $z=1.3\,\kpc$).  Under the same
interpretation \calvera's distance is $d=8.4\,{\rm kpc}$ from the
Sun, and $z=5.1\, {\rm kpc}$ from the plane of the disk, with
Galactocentric distance $R_c=14.0\,{\rm kpc}$, placing it firmly out
of the disk and in the Galactic halo.  No other isolated neutron stars
are known in the halo, although it seems likely that at least some
radio pulsars -- whose distance estimates make use of the free
electron density of the ISM, dominated by the disk component --
populate the halo \citep{manchester05}. The inferred luminosity of
\calvera\ is then $L_{\rm X}$ (0.1--2.4 keV) = 1.0\tee{34} \cgslum --
an order of magnitude greater than the next most-luminous INS
(Fig.~\ref{fig:compare}), due to the high \kteff\ of \calvera.

This location leads to a conundrum.  The interstellar medium (ISM) in
the halo is too sparse to power \calvera's \xray\ luminosity by
accretion.  If the source is then powered by remnant heat from the
supernova which produced it, then standard cooling curves require a
cooling age $\tau_{c}<\ee{6}\,{\rm yr}$ old \citep{page06} during
which time it must travel at velocity perpendicular to the plane $v_Z$
a distance of at least $z=v_z\tau_c$; its distance from the plane then
requires $v_Z > 5100$\,\kmsec, greater than has been previously
observed for any NS.  Alternatively if the INS is travelling at a more
usual $v_Z=380\,\kmsec$ \citep{faucher06}, that requires a cooling
time of $\tau_{c}>13$\,Myr, which cannot be accommodated by either
cooling of standard NSs \citep{page04} or of highly magnetized ones
\citep{arras04}.  Thus, the INS hypothesis for \calvera\ leads to
conclusions which challenge existing theory.

While high proper-motion is sometimes used to demonstrate proximity of
an astronomical object, one cannot exclude the distant,
standard-cooling INS hypothesis with a high proper-motion measurement
in this case.  This is because a more distant INS (out in the halo)
requires a proportionately greater $v_Z(\geq d\sin(b)/\tau_c)$. The
projection of this velocity across the line of sight results in a
distance-independent proper motion $\mu=\sin(b)\cos(b)/\tau_c$; for
\calvera, this is $\mu=100\, {\rm mas}\, {\rm yr}^{-1}$ for
$\tau_c$=\ee{6} yr, independent of \calvera's distance.

%%%%%%%%%%%%%%%%%%%%%%%%%%%%%%%%%%%%%%%%

\subsubsection{Magnetar}
\label{sub:magnetar}

While the observation of a slow pulse period, $P\approx 6$\,s, and a
rapid spin-down rate (for an AXP identification), or alternatively a
burst of gamma-rays (for an SGR identification), would be required to
demonstrate a magnetar nature for \calvera, neither possibility can be
ruled out with the current set of observations.  We thus consider
the hypothesis that \calvera\ is a magnetar.

Recent work has identified a tight clustering of the maximum
persistent 2--10 keV \xray\ luminosities of magnetars near
$\sim$1.3\tee{35} \cgslum\ -- making them, in a sense, \xray\ standard
candles \citep{woods06,durant06}.  Magnetar \xray\ spectra (0.1--10
keV) can be described by a soft blackbody ($\kteff=0.41$--0.63~keV)
with a power-law dominating at higher energies (power-law photon index
$\alpha=2.0$--4.5).  While the low signal to noise of our \swift\ XRT
spectrum of \calvera\ does not allow demonstration of multiple
spectral components, a single power-law fit across 0.5--10 keV is
expected to overestimate the 2--10 keV flux, and hence underestimate
the source distance according to this relation.  Our best-fit single
power-law flux (Table~\ref{tab:props}), assuming the standard-candle
luminosity derived by \cite{durant06} for \calvera, yields a distance
$d=66\kpc$, and a height above the Galactic disk of $z=40\kpc$.  This
would imply an even more extreme space velocity/lifetime (large
$v_Z\,\tau_{\rm c}$) conundrum than the INS hypothesis considered
above.

Alternatively, if (contrary to current evidence) we consider the 2--10
keV luminosity of magnetars to be a free parameter, then we note that
known magnetars \citep{durant06,muno06} are all located within
$|z|<0.1$\,kpc of the Galactic plane.  Placing \calvera\ at
$z=+0.1$\,kpc, then, yields a distance of $d<0.17\,\kpc$ and a 2--10
keV \xray\ luminosity $L_{\rm X}<$8.7\tee{29}\,\cgslum, a factor of
$10^5$ times less luminous than the standard-candle luminosity
\citep{durant06} and $10^3$ times less luminous than the next-faintest
known magnetar \citep{muno06}.

\calvera, if it is a magnetar, is thus observationally distinguished
from all other known magnetars, having either an anomalously small
luminosity or an anomalously large Galactic altitude.  We thus
consider the magnetar hypothesis strongly disfavored.  

%%%%%%%%%%%%%%%%%%%%%%%%%%%%%%%%%%%%%%%%

\subsubsection{Compact Central Object}
\label{sub:cco}

The compact central object (CCO) of Cas~A \citep{pavlov00} is one of
several point-like \xray\ sources in the centers of Galactic supernova
remnants (SNRs) that have similar spectral properties
\citep{slane99,pavlov01,park06}.  The Cas~A CCO is observed with
\kteff=$490\,{\rm eV}$ and emission radius
$\rbb=0.29\ud{0.16}{0.09}$\,km at the distance of Cas~A, $d=3.4$\,kpc.
It is unclear why CCOs have apparently smaller \rbb\ than magnetars
and INSs.  The small blackbody radii of the CCO sources are
inconsistent, in all current theoretical scenarios, with uniform
emission from the full surface of a neutron star. A small radius would
be expected if these sources emit the majority of their \xray\
radiation from hot polar caps covering a fraction of the NS surface;
however, the absence of detected pulsations from the CCOs (with the
exception of RX~J121000.8$-$522625) casts some doubt on this
interpretation.  As an alternative, it has been suggested that the
CCOs may be accreting black holes.

While the unique class-defining characteristic of a CCO is association
with a SNR, and there is no known SNR within 2 deg of \calvera, we may
consider the possibility that \calvera\ is the first ``unhosted'' CCO.

If we assume a three-dimensional space velocity for \calvera\ equal to
the $v=380\,\kmsec$ average for radio pulsars \citep{faucher06}, and
an age appropriate to the cooling age for a \ee{6} K NS,
$\tau=5\tee{5}$\,yr \citep{page04}, then we expect \calvera\ to have
traveled 190~pc from its birthplace.  Travelling at a
statistical-average angle of $30\deg$ with respect to the disk, it
would be expected to traverse a vertical distance $\delta z\approx
95$\,pc out of the Galactic plane.  Adding this distance to the
stellar scale height of the Galactic disk, $z_* \approx 90$\,pc, we
expect it to have reached $z\approx 180$\,pc by the time it has cooled
to its present temperature.  This would then yield a distance
$d=300$\,pc, an emitting radius $\rbb \approx 0.22$\,km, and a
luminosity $L_{\rm X}=1.3\tee{31} \cgslum$ (0.1--2.4 keV).  This would
make \calvera\ a factor of $\approx$10 fainter than the faintest known
CCO, while giving it a comparable temperature and emitting radius.
Such a distance is within reach of a parallax measurement with present
instruments \citep{kaplan07}.

\subsubsection{Millisecond Radio Pulsar}
\label{sub:msp}

We compare the \xray\ properties of \calvera\ to MSPs observed in the
homogeneous survey of the MSP population of 47~Tuc \citep{bogdanov06}.
\calvera\ has a similar effective temperature as these MSPs (see
Fig.~\ref{fig:compare2}); and, for distances between 80 and 260 pc,
it has a similar thermal bolometric luminosity (\ee{30-31} \cgslum;
Fig.~\ref{fig:compare}) and \rbb\ (0.06-0.18 km).  The $<$41\% r.m.s
variability limit is consistent with variability limits on these MSPs
\citep{cameron07}.  Thus, the measured properties of \calvera\ are
consistent with those of an MSP with a distance between 80 and 260
pc.

Intriguingly, one radio pulsar has previously been discovered
from a similar \xray-selected sample \citep{zepka96}.  If \calvera\ is
a radio pulsar, then, it would not be unprecedented.

% NOTE, only 0631 has an X-ray flux << Edot, so Zepka concluded the
% other two weren't associated.   No further work. 
% PSR J0631+10, J1843+20, J1908+0457; 

%%%%%%%%%%%%%%%%%%%%%%%%%%%%%%%%%%%%%%%%

%%%%%%%%%%%%%%%%%%%%%%%%%%%%%%%%%%%%%%%%

\section{Conclusions}
\label{sec:conclude}

% Summary
We have identified a new isolated compact object at high Galactic
latitude ($b=+37\deg$), \calvera, via its high \xray\ to optical flux
ratio, $F_X/F_V>8700$.

If \calvera\ is a typical INS, it would have a distance $d=11.1\,{\rm
kpc}$ which would require a spatial velocity $v_z> 6700\,\kmsec$ to
reach this location in less than a cooling time $\tau_c=1\,{\rm Myr}$.
This velocity is much greater than the $\approx 380 \kmsec$ value
typical to pulsars, and even much greater than the highest directly
measured velocity for a neutron star \citep[1083\ud{103}{90} \kmsec
,][]{chatterjee05} or the highest implied velocity \citep[800-1600
\kmsec ,][]{chatterjee04}.  Alternatively within the typical INS
hypothesis, it would have cooling time ($\tau_c\geq 17\,{\rm Myr}$)
much longer than present cooling-time predictions.

If \calvera\ is a typical magnetar, its distance would be even greater
than in the INS hypothesis, and therefore velocity and cooling-time
implications are more disparate from the present observations and
theory.

If \calvera\ is a CCO, it would avoid these velocity and cooling-time
implications, but it would be the first such object discovered outside
a SNR.

\calvera\ may be an ICBAS source (``It Can Be Any Size''), which
implicitly is not related to any of the known classes, can have any
\rbb\ and thus be at any distance and luminosity (as a function of
\rbb).  This hypothesized class of sources is simply a
counter-hypothesis to the known classes, making a point of the fact
that, in the absence of an assumed \rbb (or, equivalently, $L_X$), we
can say nothing about \calvera's distance or classification. 

The only class of objects consistent with the demonstrated properties
of \calvera\ are the radio pulsars; in particular, pulsars analogous
to the MSPs observed in the globular cluster 47~Tuc. We therefore
conclude that \calvera\ is most likely a radio pulsar.  This could be
confirmed by detection of radio pulsations from this source.

If \calvera\ is a new (perhaps fast) radio pulsar, it would be an
observationally useful object.  It exhibits an \xray\ flux equal to
the \xray-brightest and second closest millisecond radio pulsar,
PSR~J0437$-$4715 \citep{zavlin06}.  There are only 5 known radio
pulsars within $d<$260 pc, and only one in the northern hemisphere
\citep{manchester05}\footnote{http://www.atnf.csiro.au/research/pulsar/psrcat}. If
\calvera\ indeed turns out to be an MSP, it would be the third closest
MSP in the sky (after PSR~J0437-4715 at $d=160$ pc and PSR~J2124-3358
and $d=250$ pc) and potentially the closest MSP at $d=80$ pc.  It
would be the closest MSP in the northern hemisphere (followed by
PSR~J0030+0451, d=300 pc) making it a potentially useful target both
for a pulsar timing array \citep{pta} and for targeted search with the
{\em Laser Interferometric Gravitational Wave Observatory} (LIGO;
e.g. \citealt{cutler02}).

%%%%%%%%%%%%%%%%%%%%%%%%%%%%%%%%%%%%%%%%

\acknowledgements

We would like to express our gratitude to the \swift\ operations team
for the fill-in target observations that made this work possible. We
gratefully acknowledge Jean-Rene Roy for granting director's
discretionary time with Gemini-North for this project.  We are also
grateful to Harvey Tananbaum for granting director's discretionary
time with \chandra\ for this project.  We gratefully acknowledge and
thank the anonymous referee, for suggestions regarding the SWIFT data
analysis which corrected an earlier version, changing the quantitative
conclusions.  RER gratefully acknowledges useful conversations with
A. Cumming, an advocate of the strong ICBAS hypothesis.  RER is
supported by the Natural Sciences and Engineering Research Council of
Canada Discovery program.

%\bibliographystyle{apj_8}
%\bibliography{journals_apj,magrefs,bobs}

%%%%%%%%%%%%%%%%%%%%%%%%%%%%%%%%%%%%%%%%

\clearpage

%%%%%%%%%%%%%%%%%%%%%%%%%%%%%%%%%%%%%%%%%%%%%%%%%%%%%%%%%%%%%%%%%%%%%%

%%%%%%%%%%%%%%%%%%%%%%%%%%%%%%

\clearpage

\newpage

%%%%%%%%%%%%%%%%%%%%%%%%%%%%%%
\begin{deluxetable}{ll}
\tablewidth{12cm}
\tablecaption{Characteristics of \calvera \label{tab:props} }
\tablehead{
\colhead{Characteristic}  & 
\colhead{Value} 
}
\startdata
Right Ascension (J2000)   & $14^{\rm h}12^{\rm m}55\fs 885$ \\ 
Declination (J2000)       & $+79\deg 22\arcmin 04\farcs 10$ \\ 
Uncertainty radius (90\%)  & 0.57\arcsec \\
UVOT Limit                & $f_{\rm UVM2} < 1.3\times 10^{-17}$~\cgsflux \\
Gemini Limit (3$\sigma$)  & $g > 26.3$\,mag \\
\hline
\multicolumn{2}{c}{Blackbody Energy Spectrum}\\
\hline
\kteff              &  215\ppm 25  eV  \\ 
%Normalization       & 5.4\ud{2.7}{1.9} ($R_{\rm km}/D_{10 {\rm kpc}})$ \\ 
Normalization       & 7.2\ud{2.4}{1.8} ($R_{\rm km}/D_{10 {\rm kpc}})$ \\ 
%Corrected X-ray Flux  & 7.6\tee{-13} (\cgsflux; 0.1--2.4 keV) \\ 
Corrected X-ray Flux  & 1.2\tee{-12} (\cgsflux; 0.1--2.4 keV) \\ 
$N_H$ (fixed)       & 3\tee{20} \perval{cm}{-2} \\
C-statistic       & 23.97 \\ 
\hline
\multicolumn{2}{c}{Power Law Energy Spectrum}\\
\hline
Photon Slope $\alpha$       & 2.8\ppm0.3 \\
Corrected X-ray Flux        &  2.5\tee{-13} (\cgsflux; 2-10 keV) \\
$N_H$ (fixed)               &  3\tee{20} \perval{cm}{-2}     \\
C-statistic  & 30.03 \\
\hline
\enddata
\tablecomments{Properties of the isolated neutron star \calvera\
(\rxsj). Parameters are estimated using the \xray\ spectral modelling
package XSPEC v12.3.0 \citep{xspec}.  The \xray\ equivalent hydrogen
column density was fixed at the value of the integrated column through
the galaxy in the direction of the source
\citep{dickey90}. Uncertainties are quoted at 90\% confidence.  \xray\
fluxes are corrected for interstellar absorption. Blackbody and
power-law spectral parameters are independently derived by fitting to
the same \swift\ XRT data, over 0.5--10\,keV.
}
\end{deluxetable}
%%%%%%%%%%%%%%%%%%%%%%%%%%%%%%

%%%%%%%%%%%%%%%%%%%%%%%%%%%%%%

\clearpage
\begin{deluxetable}{rrrrrrrrrr}
\tabletypesize{\scriptsize}
\tablewidth{0pt}
\tablecaption{ \label{tab:dist} Galacto-centric Positions of INSs and
  \calvera\ in an INS Interpretation}
\tablehead{
\colhead{} & 
\colhead{\kteff} & 
\colhead{$F_X$} & 
\colhead{(l,b)} & 
\colhead{$X$} & 
\colhead{$Y$} & 
\colhead{$Z$} & 
\colhead{d} & 
\colhead{$R_c$} & 
\colhead{} \\
\colhead{Source} & 
\colhead{(eV)} & 
\colhead{} & 
\colhead{(deg,deg)} & 
\colhead{(kpc)} & 
\colhead{(kpc)} & 
\colhead{(kpc)} & 
\colhead{(kpc)} & 
\colhead{(kpc)} & 
\colhead{Refs.} 
}
\startdata
  1RXS~J0420.0$-$5022&   45 &   5      &   258, -44 & -0.36 & 8.58 & -0.35 & 0.51  &  8.59 & 1  \\
     RXJ0720.4$-$3125&   90 &   100    &   244, -8  & -0.45 & 8.72 & -0.07 & 0.50  &  8.73 & 2  \\
     RXJ0806.4$-$4123&   95 &   2.8    &   257, -5  & -3.29 & 9.26 & -0.30 & 3.39  &  9.83 & 3  \\
1RXS~J130848.6+212708&  117 &   45     &   339, 83  & -0.06 & 8.35 &  1.29 & 1.30  &  8.45 & 4  \\
          \calvera   &  215 &   12     &   118,  37 &  5.9  & 11.66&  5.08 & 8.43 &  14.04& present \\
    1RXS~J1605.3+3249&   91 &   88     &    53,  48 &  0.30 & 8.27 &  0.42 & 0.56  &   8.29& 5 \\
\eighteen\           &   63.5 & 210    &   359, -17 &  0.00 & 8.34 & -0.05 & 0.167 &   8.34& 6 \\
1RXS~J214303.7+065419&   91 &   87     &    63, -33 &  0.42 & 8.29 & -0.31 & 0.56  &   8.30& 7\\
\enddata
\tablecomments{Galactic positions of the seven INSs, plus \calvera,  under
  the assumption all have the same \rbb\ as \eighteen\ at a distance
  of 167 pc (see text). Reading across the columns, we
give the source name, the measured effective temperature, the X-ray
flux in units of \ee{-13} \cgsflux\ ($0.1-2.4\,{\rm keV}$); the
galactic longitude and latitude (l,b); the resulting galactic three
dimensional coordinates $X$, $Y$, and $Z$, where (0,0,0) is Galactic
Center, and (0,8.5,0) is the Sun's location \citep{taylor93}; the
source's distance from the Sun $d$; and galacto-centric distance
$R_c$, with the relevant references.  These positions are plotted in
Fig.~\ref{fig:dist}. }
\tablerefs{
1,    \cite{haberl04};	      
2,     \cite{haberl06a};
3,     \cite{haberl04};
4,     \cite{schwope99};
5,      \cite{motch99};
6,      \cite{burwitz03,kaplan07};
7,      \cite{zampieri01}
}  
\end{deluxetable}

%%%%%%%%%%%%%%%%%%%%%%%%%%%%%%%%%%%%%%%%%%%%%%%%%%%%%%%%%%%%%%%%%%%%%%
%%%%%%%%%%%%%%%%%%%%%%%%%%%%%%%%%%%%%%%%%%%%%%%%%%%%%%%%%%%%%%%%%%%%%%
%%%%%%%%%%%%%%%%%%%%%%%%%%%%%%%%%%%%%%%%%%%%%%%%%%%%%%%%%%%%%%%%%%%%%%

\end{document}